\documentclass[prb,
12pt,
superscriptaddress,showpacs,amsmath,amssymb]{revtex4}
\usepackage{amsfonts}
\usepackage{bm}
\usepackage{verbatim}

\usepackage{graphicx}

\begin{document}

\author{G.E.~Volovik}
\affiliation{Low Temperature Laboratory, Aalto University,  P.O. Box 15100, FI-00076 Aalto, Finland}
\affiliation{Landau Institute for Theoretical Physics, acad. Semyonov av., 1a, 142432,
Chernogolovka, Russia}

\title{Discrete $Z_4$ symmetry in quantum gravity}

\date{\today}

\begin{abstract}
We consider the discrete $Z_4$ symmetry ${\hat {\bf i}}$, which takes place in the scenario of quantum gravity where the gravitational tetrads emerge as the order parameter -- the vacuum expectation value of the bilinear combination of fermionic operators. Under this symmetry operation $\hat {\bf i}$, the emerging tetrads are multiplied by the imaginary unit, $\hat {\bf i}\,e^a_\mu = -i e^a_\mu$. The existence of such symmetry and the spontaneous breaking of this symmetry are also supported by the consideration of the symmetry breaking scheme in the topological superfluid $^3$He-B. The order parameter in $^3$He-B is also  the bilinear combination of the fermionic operators. This order parameter is the analogue of the tetrad field, but it has complex values. The $\hat {\bf i}$-symmetry operation changes the phase of the complex order parameter by $\pi/2$, which corresponds to the $Z_4$ discrete symmetry in quantum gravity. We also considered the alternative scenario of the breaking of this $Z_4$ symmetry, in which the $\hat {\bf i}$-operation changes sign of the scalar curvature, $\hat {\bf i} \,{\cal R} = - {\cal R}$, and thus the Einstein-Hilbert action violates the $\hat {\bf i}$-symmetry. In the alternative scenario of symmetry breaking, the gravitational coupling $K=1/16\pi G$ plays the role of the order parameter, which changes sign under $\hat {\bf i}$-transformation.
\end{abstract}

\maketitle

\newpage
\tableofcontents

\section{Introduction}

The discrete symmetries, such as $P$, $T$ and $C$ symmetries, play an important role in particle physics,  in gravity and cosmology,\cite{Boyle2018} as well as in topological matter.\cite{Ryu2016}
We consider the discrete $Z_4$ symmetry ${\hat {\bf i}}$, which takes place in the Akama-Diakonov-Wetterich (ADW) scenario of quantum gravity. In this scenario, the gravitational tetrads emerge as the symmetry breaking order parameter -- the vacuum expectation values of the bilinear combinations of the fermionic operators.\cite{Akama1978,Wetterich2004,Wetterich2022,Diakonov2011,VladimirovDiakonov2012,VladimirovDiakonov2014,ObukhovHehl2012,Maiezza2022}
 Under the symmetry operation $\hat {\bf i}$, the tetrads are multiplied by the imaginary unit, $\hat {\bf i}\,e^a_\mu = -i e^a_\mu$. The similar symmetry breaking scenario characterizes the topological superfluid $^3$He-B. But in  $^3$He-B, the order parameter, which corresponds to tetrads in quantum gravity, is the complex matrix. 
The $\hat {\bf i}$-symmetry operation changes the phase of this complex order parameter by $\pi/2$, which corresponds to the $Z_4$ discrete symmetry in quantum gravity.

In Sec. \ref{composite} we consider the  $\hat {\bf i}$-symmetry in the ADW scenario. The  consequences of the spontaneous breaking of $\hat {\bf i}$-symmetry is discussed in Sec. \ref{Brokeni}. The Sec. \ref{alternative} is devoted to the alternative scenario of the symmetry breaking, in which the scalar curvature is not invariant under the $\hat {\bf i}$-operation:  it changes sign, $\hat {\bf i} {\cal R} = - {\cal R}$.

\section{Composite tetrads and ${\hat {\bf i}}$-symmetry}
\label{composite}

In the Akama-Diakonov-Wetterich (ADW) approach, the gravitational tetrads appear as composite objects made of the more fundamental fields, the quantum fermionic fields: \cite{Akama1978,Wetterich2004,Wetterich2022,Diakonov2011,VladimirovDiakonov2012,VladimirovDiakonov2014,ObukhovHehl2012,Maiezza2022}
\begin{equation}
 \hat E^a_\mu = \frac{1}{2}\left( \Psi^\dagger \gamma^a\partial_\mu  \Psi -  \Psi^\dagger\overleftarrow{\partial_\mu}  \gamma^a\Psi\right) \,.
\label{TetradsFermionsOperators}
\end{equation}
The original action does not depend on tetrads and metric and is described solely in terms of differential forms:
\begin{equation}
S=\frac{1}{24}e^{\alpha\beta\mu\nu} e_{abcd} \int d^4x  \, \hat E^a_\alpha   \hat E^b_\beta \hat E^c_\mu \hat E^d_\nu \,.
\label{OriginalActions}
\end{equation}
This operator analog of the cosmological term has high symmetry. It is symmetric under coordinate transformations $x^\mu \rightarrow \tilde x^\mu(x)$, and thus is also scale invariant. In addition, the action is symmetric under spin rotations, or under the corresponding gauge transformations when the spin connection is added to the gradients. 

For us it is important that this action is also symmetric under the complex coordinate transformation $x^\mu \rightarrow i x^\mu(x)$. Let us denote this symmetry as ${\hat {\bf i}}$-symmetry:
\begin{equation}
{\hat {\bf i}}\, x^\mu = i x^\mu \,,
\label{iSymmetry}
\end{equation}
Since the field operator $\hat E^a_\mu$ in Eq.(\ref{TetradsFermionsOperators}) is linear in gradients and represents the 1-form, ${\bf E}={\bf E}_\mu dx^\mu$,  it is multiplied by $-i$ under this symmetry operation:
\begin{equation}
{\hat {\bf i}}\, \hat E^a_\mu =  -i\hat E^a_\mu \,.
\label{TetradsTransformation}
\end{equation}
The operator $\hat E^a_\mu$ is Hermitian or anti-Hermitian, once representation of 
$\gamma$-matrices is specified. Under the symmetry transformation 
(\ref{iSymmetry}) the operator $\hat E^a_\mu$ becomes correspondingly anti-Hermitian or Hermitian, but the action (\ref{OriginalActions}) remains invariant under this ${\hat {\bf i}}$-transformation.

The action may also contain the operator analog of the Einstein–Hilbert–Cartan term\cite{Diakonov2011}, 
\begin{equation}
e^{\alpha\beta\mu\nu} e_{abcd} \int d^4x  \,\hat E^a_\alpha   \hat E^b_\beta \hat F^{cd}_{\mu\nu}  \,.
\label{Cartan}
\end{equation}
Here 
$\hat F^{cd}_{\mu\nu}$ is the operator of the Cartan curvature 2-form. As the 2-form, 
${\bf F}={\bf F}_{\mu\nu} dx^\mu \wedge dx^\nu$, it changes sign under ${\hat {\bf i}}$-transformation:
\begin{equation}
{\hat {\bf i}}\,\hat F^{cd}_{\mu\nu} = -  \hat F^{cd}_{\mu\nu} \,,
\label{CurvatureTransformation}
\end{equation}
while the action (\ref{Cartan}) is  ${\hat {\bf i}}$-invariant.

\section{Broken ${\hat {\bf i}}$-symmetry}
\label{Brokeni}

In the ADW scenario of quantum gravity, the tetrads $e^a_\mu$ emerge as the order parameter of the spontaneous symmetry breaking. These are the vacuum expectation values of the bilinear fermionic 1-form $\hat E^a_\mu$:
\begin{equation}
e^a_\mu=<\hat E^a_\mu>\,.
\label{TetradsFermions}
\end{equation}
If we use the Hermitian choice of the operator $\hat E^a_\mu$, the emergent tetrads $e^a_\mu$ become the real functions. Since in the ADW theory the fermionic fields are dimensionless, \cite{VladimirovDiakonov2012}
the covariant tetrads have dimension of the inverse length, $[e^a_\mu]=1/[length]$.
 
The tetrad order parameter breaks the separate symmetries under orbital and spin transformations, but remains invariant under the combined rotations. On the level of the Lorentz symmetries the symmetry breaking scheme is $L_L \times L_S \rightarrow L_J$. Here $L_L $ is the group of Lorentz transformations in the coordinates space, $L_S $ is the group of Lorentz transformations in the spin space, and 
$L_J$ is the residual symmetry. It is the symmetry group of the order parameter, which is invariant under the combined Lorentz transformations $L_J$. Note also that the discrete $P$ and $T$ symmetries of the Standard Model  are also the combined symmetries, since they include both the coordinate transformations and transformations of the fermionic fields.

In addition, the order parameter in Eq.(\ref{TetradsFermions}) breaks the discrete ${\hat {\bf i}}$-symmetry and becomes anti-Hermitian under ${\hat {\bf i}}$-transformation:
\begin{equation}
{\hat {\bf i}}\,e^a_\mu = -i e^a_\mu\,.
\label{TetradsSymmetry2}
\end{equation}
We can compare this symmetry breaking with the spontaneous breaking of the $PT$-symmetry proposed in Ref. \cite{Vergeles2023}, where the special type of the $PT$-symmetry operation changes sign of all tetrads, $PT e^a_\mu =- e^a_\mu$. In connection to the transformation of tetrads, the ${\hat {\bf i}}$-symmetry corresponds to the square root of this $PT$-symmetry: 
\begin{equation}
{\hat {\bf i}}^2 \, e^a_\mu=PT \, e^a_\mu = -e^a_\mu
\,.
\label{TetradsSymmetryPT}
\end{equation}
That is why the ${\hat {\bf i}}$-operation belongs to the discrete $Z_4$ group, $({\hat {\bf i}}, {\hat {\bf i}}^2=PT, {\hat {\bf i}}^3= - {\hat {\bf i}}, {\hat {\bf i}}^4=1)$.

The ADW symmetry breaking mechanism of emergent gravity has the analogue in the spin-triplet $p$-wave superfluids, where the effective gravitational vielbein also emerge as the bilinear fermionic 1-form \cite{Volovik1990,Volovik2022}. In the B-phase of superfluid $^3$He the symmetry breaking scheme is $SO(3)_L \times SO(3)_S \rightarrow SO(3)_J$, where $SO(3)_L$ and $SO(3)_S$ are correspondinglly the orbital and spin rotation groups, and $SO(3)_J$ is the residual symmetry -- the symmetry of the order parameter under combined rotations. Here ${\bf J}$ is the total angular momentum operator ${\bf J}={\bf L}+{\bf S}$.
Such symmetry breaking to the diagonal subgroup is known in superfluid $^3$He as the relative symmetry breaking.\cite{Leggett1973} This means that the  symmetry under the separate rotations in spin and orbital spaces is broken, while the properties of $^3$He-B are isotropic due to the symmetry under the combined rotations. 

It is important, that in $^3$He-B, the $U(1)$ symmetry is also spontaneously broken. Under the $U(1)$ phase rotations, the order parameter (the triad analog of tetrads) is transformed according to $e^a_\mu \rightarrow e^{i \phi} e^a_\mu$. Then the ${\hat {\bf i}}$-symmetry operation in Eq. (\ref{TetradsSymmetry2}) corresponds in $^3$He-B to the global $U(1)$ transformation with the phase $\phi=-\pi/2$. This again demonstrates that with respect to tetrads, the  ${\hat {\bf i}}$-symmetry is the element of the discrete $Z_4$-symmetry group of quantum gravity with $\phi=\pi n/2$. , which extends the symmetry group. The possible extension of this $Z_4$ group to the full $U(1)$ group of quantum mechanics is discussed in Sec. \ref{imaginary} in connection with another phase of superfluid $^3$He, the the planar phase with massless Dirac fermions.

Note, that in our case the symmetry operation ${\hat {\bf i}}$ does not include the transformations of the fermionic and bosonic operators and the interchanges of the Dirac variables with their Hermitian conjugates.
It is the pure coordinate transformation of the operators, ${\hat {\bf i}} \Psi(x^\mu)=\Psi(ix^\mu)$, i.e. these operators are the world scalars under such discrete kind of diffeomorphisms.

\subsection{Metric as fermionic quartet}

In the broken symmetry state in the ADW scenario, the metric field emerges as the secondary object -- the bilinear combination of the tetrad fields:
 \begin{equation}
g_{\mu\nu}=\eta_{ab}e^a_\mu e^b_\nu \,.
\label{DimMetric}
\end{equation}
That is why, in this quantum gravity the metric is the fermionic quartet.
Under the ${\hat {\bf i}}$-transformation the signature of the metric changes:
 \begin{equation}
{\hat {\bf i}}\,g_{\mu\nu} = - g_{\mu\nu} \,.
\label{MetricSymmetry}
\end{equation}

Let us mention the connection with the scenario suggested in Refs. \cite{BondarenkoZubkov2022,Bondarenko2022}, where the signature of the metric is represented by the dynamical variable $O_{ab}$. Then the tensor $\eta_{ab}$ in Eq. (\ref{DimMetric}) emerges as the vacuum expectation value $\eta_{ab}=<O_{ab}>$ in the corresponding symmetry breaking phase transition.   

Let us also mention the complexification of the tetrads\cite{Wetterich2022} and the complexification of the Lorentz group in Refs. \cite{Zlosnik2024a,Zlosnik2024b} and references therein.  The scenario of emergence of gravity from the breaking of gauge symmetry in Refs. \cite{Zlosnik2024a,Zlosnik2024b} is similar to the scenario with the so-called translational gauge fields in crystals, where the gravitational tetrads emerge from the elasticity tetrads describing the elasticity theory in crystals.\cite{DzyalVol1980,NissinenVolovik2019,NissinenVolovik2018,KlinkhamerVolovik2019,Burkov2023} 
 In the elasticity theory, an arbitrarily deformed crystal structure is described as a system of the crystallographic surfaces of constant phase $X^a(x)=2\pi n^a$, and the elasticity tetrads are $e^{a}_\mu = \partial_\mu X^a$. Note, that the elasticity tetrads change sign under the conventional $PT$ transformation of coordinates, and thus the 
 $\hat {\bf i}$-symmetry is the square root of this $PT$ symmetry.
 
 In principle, the so-called vestigial gravity is possible in quantum gravity, where  the tetrad order parameter is absent, $e_\mu^a=< \hat E^a_\mu> =0$, while the metric emerges as the vacuum expectation value of the bilinear combination of tetrad operators, $g_{\mu\nu} =\eta_{ab}< \hat E^a_\mu \hat E^b_\nu>$.\cite{Volovik2024g} In the vestigial gravity the Equivalence Principle is violated, since such gravity acts with different strength on fermions and bosons, and they do not follow the same trajectories in a given gravitational field. 

\subsection{Interval and scalar field}

While the metric in Eq.(\ref{MetricSymmetry}) changes sign under the ${\hat {\bf i}}$-transformation, the interval remains ${\hat {\bf i}}$-invariant due to Eqs. (\ref{iSymmetry}) and (\ref{MetricSymmetry}):
 \begin{equation}
ds^2=g_{\mu\nu}dx^\mu dx^\nu \,\,, \,\,   \, {\hat {\bf i}} \,ds^2 = ds^2\,.
\label{DimensionInterval}
\end{equation}
This makes ${\hat {\bf i}}$-invariant the classical action $S=M\int ds$ for massive particle, ${\hat {\bf i}} \,S=S$.

The emergence of the metric gives rise to the quadratic action for the scalar field $\Phi$:\begin{equation}
S=\int d^4 x\,\sqrt{-g} \, g^{\mu\nu} \nabla_\mu \Phi^*  \nabla_\nu \Phi\,,
\label{scalar}
\end{equation}
which is also ${\hat {\bf i}}$-invariant. This is because the coordinate transformation of the gradients of the scalar field is compensated by the change of the sign of the metric:
$\hat {\bf i}\, \nabla_\mu = -i \nabla_\mu$,  $\hat {\bf i} \, g^{\mu\nu} = - g^{\mu\nu}$ and $\hat {\bf i} \, g = g$.

\subsection{Gauge fields}

The gauge potential is the 1-form field, $A=A_\mu dx^\mu$, and thus it transforms in the same way as gradient:
 \begin{equation}
{\hat {\bf i}}\, A_\mu = - i A_\mu\,.
\label{PotentialSymmetry}
\end{equation}
The zero-form action describing the interaction of a charged point particle with the $U(1)$ gauge field remains invariant under the ${\hat {\bf i}}$-transformation:
 \begin{equation}
S=q \int dx^\mu A_\mu  \,\,, \,\, {\hat {\bf i}}\, S=S\,,
\label{ChargeParticleAction}
\end{equation}
where $q$ is the dimensionless electric charge.
On the other hand, the field strength, being the 2-form, $F=F_{\mu\nu} dx^\mu \wedge dx^\nu$, changes sign under the ${\hat {\bf i}}$-transformation:
 \begin{equation}
{\hat {\bf i}}\,  F_{\mu\nu} = - F_{\mu\nu}\,.
\label{FieldSymmetry}
\end{equation}
The quadratic action for the gauge field, which appears after the emergence of the metric, is quadratic in $F_{\mu\nu}$ and is quadratic in $g^{\mu\nu}$:
\begin{equation}
S \propto \int d^4 x\,\sqrt{-g} \,F_{\mu\nu} F^{\mu\nu} \,,
\label{GaugeAction}
\end{equation}
 and thus it is ${\hat {\bf i}}$-invariant.

\subsection{Fermions}

The action for the massive Dirac particles 
\begin{equation}
S=\int d^4x\,  e\, (ie^\mu_a \bar\Psi  \gamma^a \nabla_\mu \Psi - M\bar\Psi \Psi)\,.
\label{Fermions}
\end{equation}
is also ${\hat {\bf i}}$-invariant. This is because of the following transformations of the contravariant tetrads, gradients and the tetrad determinant:
\begin{equation}
{\hat {\bf i}}\,  e^\mu_a = i e^\mu_a  \,\,,\,\, {\hat {\bf i}}\,  \nabla_\mu = -i \nabla_\mu  \,\,,\,\, {\hat {\bf i}}\,  e = e\,.
\label{FermionsSymmetry}
\end{equation}
Let us recall, that the symmetry operation ${\hat {\bf i}}$ is the pure coordinate transformation, which does not involve the $\gamma$-matrices. Note also that in the ADW theory the contravariant tetrads have dimension of the length, $[e_a^\mu]=[length]$, while the fermionic fields and the mass $M$ are dimensionless.\cite{VladimirovDiakonov2012,Volovik2022g} That is why the action (\ref{Fermions}) is dimensionless.

 \subsection{Gravity from bosons}
 
 Till now we considered gravity emerging in the background of the fermionic vacuum. The more restricted gravity emerges in the bosonic background, see e.g. Ref. \cite{Goulart2011}. Now, instead of tetrads the metric field is the emerging gravitational variable. Example of the corresponding order parameter as the vacuum expectation value of the bosonic fields is:
 \begin{equation}
g_{\mu\nu} \propto \left<\nabla_\mu \Phi^\dagger \nabla_\nu \Phi +\nabla_\nu \Phi^\dagger \nabla_\mu \Phi\right>\,.
\label{BosonicMetricEq}
\end{equation}
Under ${\hat {\bf i}}$-transformation the effective metric changes sign in the same way as in the fermionic vacuum, ${\hat {\bf i}}g_{\mu\nu} =-g_{\mu\nu}$. That is why the quadratic action for the scalar field in Eq.(\ref{scalar}) remains ${\hat {\bf i}}$-invariant.
 
  \subsection{Discrete $Z_4$ symmetry and imaginary unit in quantum mechanics}
  \label{imaginary}
 
The complexification of the coordinate transformations using $x^\mu \rightarrow i x^\mu$ becomes more transparent when complex numbers are expressed in terms of real numbers. The latter also explains why only the elements of the discrete subgroup $Z_4$ of the $U(1)$ symmetry group participates in the coordinate transformations. 

This may also have some connection to the fundamental problem of the role of the complex numbers in quantum mechanics. As is known, Schrödinger strongly resisted to introduce such product of human mind as $\sqrt{-1}$  into the wave equations.\cite{Yang2003} The possible solution of the problem is to express the effective imaginary unit in terms of the real $2\times 2$ matrix:
 \begin{equation}
a+ib  \equiv  \begin{pmatrix} a\\
b
\end{pmatrix}
\equiv a \hat {\bf I}  + b  \hat {\bf i}\,,
\label{unity}
\end{equation}
where
 \begin{equation}
\hat {\bf I}  =
 \begin{pmatrix}1& 0\\
0&1
\end{pmatrix}
 \,\,,\,\, 
\hat {\bf i} =\begin{pmatrix}0& 1\\
-1&0
\end{pmatrix} \,\,,\,\, \hat {\bf i}^2=-1 ~.
\label{imaginary_unit}
\end{equation}

The possible topological origin of the emergence of these real matrices in quantum mechanics is discussed in Refs. \cite{VolovikZubkov2014a,VolovikZubkov2014b}. The main role in this scenario, which gives rise to the effective imaginary unit, is played by the topology of exceptional points of the level crossing in the fermionic spectrum -- the so-called conical, diabolic, Dirac and Weyl points. 

Another possible origin of the effective imaginary unit $i_{\rm eff}$ is discussed by Adler in the theory of the trace dynamics.\cite{Adler2023}  According to Adler\cite{Adler2023} the emergent quantum theory may have two sectors, one with imaginary unit $i$ and one with imaginary unit $-i$. This corresponds to the $Z_2$ symmetry between the states with:  
\begin{equation}
\hat {\bf i} =\pm\begin{pmatrix}0& 1\\
-1&0
\end{pmatrix} \,.
\label{sign_i}
\end{equation}
This can be responsible for the dark matter arising in the hidden sector of the standard model with the opposite $i$.\cite{Adler2023b}

The full $Z_4$ group $(\hat {\bf I}, \hat {\bf i}, \hat {\bf i}^2, \hat {\bf i}^3)$  for the considered coordinate transformations comes from the symmetry between the Hermitian and anti-Hermitian presentations of the coordinate and momentum operators, which are similar to the Hermitian and anti-Hermitiian representations of $\gamma$-matrices in Sec. \ref{composite}.
The Hermitian matrices for the real valued coordinates and momenta are:
 \begin{equation}
\hat {\bf x} = 
\begin{pmatrix}
x& 0\\
0&x
\end{pmatrix}
\,\,,\,\,
\hat {\bf p} =
 \begin{pmatrix}
 0&- \partial_x\\
\partial_x&0
\end{pmatrix} \,\,,\,\,
\hat {\bf x} \hat {\bf p}-\hat {\bf p} \hat {\bf x}  = \hat {\bf i} \,.
\label{xdx}
\end{equation}

 The anti-Hermitian presentation of the coordinate and momentum operators, which is obtained by the $\hat {\bf i}$ element of the $Z_4$ symmetry group, $x^\mu \rightarrow ix^\mu$, can be written also in terms of real numbers:
 \begin{equation}
 \hat {\bf x}_{\rm anti}=\hat {\bf i} \, \hat {\bf x}=
\begin{pmatrix}0& x\\
-x&0
\end{pmatrix}\,\,,\,\,
\hat {\bf p}_{\rm anti}=-\hat {\bf i} \,\hat {\bf p}=-\begin{pmatrix} \partial_x &0\\
0& \partial_x 
\end{pmatrix} \,,
\label{anti_xdx}
\end{equation}
with the same canonical commutations relation as in Eq.(\ref{xdx}).

 It is important that in some cases the discrete symmetry can be automatically extended to the continuous symmetry, as it happens for the planar phase of superfluid $^3$He, where $Z_2$ symmetry is extended to $U(1)$.
 \cite{Makhlin2014}
In the planar phase, the $Z_2$ symmetry $C$ is the
combination of the spin $\pi$ rotation about the $z$ axis and
the phase rotation by $\pi/2$. In the linear approximation, the single-particle
Hamiltonian and Green function commute not only with $C$ but also with the full $SO(2)\equiv U(1)$ group of
transformations  $\exp(i\alpha C)$ generated by $C$. Since quantum mechanics is the linear theory, one may also expect that in the same way the $U(1)$ transformation $e^{i\alpha}$ of the wave function in quantum mechanics emerges as the extension of the $Z_4$ symmetry group with its discrete phases, $\alpha=n\pi/2$.

 \section{Alternative broken symmetry}
 \label{alternative}

 It can be interesting to consider the other possible scenarios of quantum gravity with different schemes of the breaking of the ${\hat {\bf i}}$-symmetry. Let us consider as an example the symmetry breaking scheme in which the metric remains invariant under the ${\hat {\bf i}}$-transformation of coordinates, $x^\mu \rightarrow ix^\mu$:
  \begin{equation}
{\hat {\bf i}}\,g_{\mu\nu} = g_{\mu\nu} \,.
\label{MetricSymmetry2}
\end{equation}
Such gravity does not include fermions, and concerns only the bosonic scalar and gauge fields interacting with the gravitational field.

\subsection{Scalar field}

 The action for the scalar field is modified, since the Eq.(\ref{scalar}) is quadratic in gradients and thus is not ${\hat {\bf i}}$-invariant. But the 4-th order gradient terms in action are invariant under ${\hat {\bf i}}$-transformation, examples of such terms are:
 \begin{equation}
S=\int d^4 x\,\sqrt{-g} \,\left( a_4 g^{\mu\nu}  g^{\alpha\beta} \nabla_\mu \nabla_\nu \Phi^*  \nabla_\alpha \nabla_\beta \Phi + b_4 \left(g^{\mu\nu} \nabla_\mu \Phi^*  \nabla_\nu \Phi \right)^2 \right)\,.
\label{scalarAlternative}
\end{equation}
 The  2-nd order gradient term, which violates the ${\hat {\bf i}}$-symmetry, may appear only in the state with the spontaneously broken ${\hat {\bf i}}$-symmetry. The corresponding order parameter can be, for example, the following vacuum expectation value of the bosonic operators: 
  \begin{equation}
 \lambda= < g^{\mu\nu} \nabla_\mu \Phi^\dagger  \nabla_\nu \Phi> \,.
\label{OPAlternative}
\end{equation}
This gives the following 2-nd order gradient term: 
  \begin{equation}
S_2=2b_4 \lambda\int d^4 x\,\sqrt{-g} \,g^{\mu\nu} \nabla_\mu \Phi^*  \nabla_\nu \Phi\,.
\label{QuadraticAlternative}
\end{equation}
The main difference from the Eq.(\ref{scalar}) is that the Eq.(\ref{QuadraticAlternative}) contains the parameter $\lambda$, which changes sign under ${\hat {\bf i}}$-transformation,
${\hat {\bf i}}\,\lambda=-\lambda$, and thus the scalar field is transformed to the ghost field.
 
 \subsection{Gauge field}
 
 In this scenario of symmetry breaking, the Eq.(\ref{FieldSymmetry}) for the ${\hat {\bf i}}$-transformation of the gauge field remains the same, since the 2-form field does not depend on metric:
  \begin{equation}
{\hat {\bf i}}\,  F_{\mu\nu} = - F_{\mu\nu}\,.
\label{FieldSymmetry2}
\end{equation}
The action for the gauge field in Eq.(\ref{GaugeAction}) is quadratic in the metric field and thus remains ${\hat {\bf i}}$-invariant. This actually follows from the scale invariance of this action.
 \subsection{Gravity}

 In the scenario of the broken symmetry in Sec. \ref{Brokeni}, where the tetrads serve as the order parameter of the symmetry breaking, the term in Eq. (\ref{Cartan}) is reduced to the conventional Einstein-Hilbert action
 $\int d^4x\sqrt{-g} K{\cal R}$. It is ${\hat {\bf i}}$-invariant, since ${\hat {\bf i}} \, {\cal R}={\cal R}$. 
  In the alternative symmetry breaking scenario discussed here, the curvature ${\cal R}$ is not ${\hat {\bf i}}$-invariant, since the transformation $x^\mu \rightarrow i x^\mu$ is not compensated by the transformation of the metric, and one obtains ${\hat {\bf i}}\,{\cal R} = - {\cal R}$. As a result, the Einstein-Hilbert action is not 
 ${\hat {\bf i}}$-invariant, and the  ${\hat {\bf i}}$-invariant gravity is the gravity, which is quadratic in the spacetime curvature. 
 
 The ${\hat {\bf i}}$-invariance of the Einstein-Hilbert action is restored if the gravitational coupling changes sign under this transformation, ${\hat {\bf i}}\,K = -K$. This means that in this scenario the gravitational coupling $K$ either serves as the order parameter of such symmetry breaking or is proportional to the order parameter $\lambda$  in Eq.(\ref{OPAlternative}), $K\propto \lambda$. In this sense, the  ${\hat {\bf i}}$-symmetry has much in common with the scale invariance.

 \section{Discussion}
 \label{DiscussionSec} 

In topological superfluid $^3$He-B, the symmetry breaking scheme includes the reduction $SO(3)_L \times SO(3)_S \rightarrow
SO(3)_J$. Here $SO(3)_L$ is the group of oribtal rotations; $SO(3)_S$ is the group of spin rotations, and $SO(3)_J$ is the residual symmetry -- the symmetry under combined rotations. 

In the ADW scenario of quantum gravity, there is the similar symmetry breaking scheme, but in the form of the Lorentz symmetries: $L_L \times L_S \rightarrow L_J$. Here $L_L =S(3,1)_L$ is the group of Lorentz transformations in the coordinates space, $L_S=S(3,1)_S$ is the group of Lorentz transformations in the spin space, and 
$L_J =S(3,1)_J$ is the residual symmetry -- the symmetry group of the order parameter. The order parameter here is the tetrad field $e^a_\mu$, which is not invariant under separate Lorentz transformations, but is invariant under the residual symmetry group -- the combined Lorentz transformations $L_J$. Two Lorentz symmetry groups as independent transformations of coordinates and spins have been also considered in Ref. \cite{Cahill2024}.

The difference between the $^3$He-B scenario and the ADW scenario of quantum gravity is not only in the different dimensions: we have 3D space dimension in $^3$He-B and (3+1) dimension of space time in the ADW gravity, and as a result, the order parameter in  $^3$He-B represents the triad field instead of tetrads.
There is another important difference: in $^3$He-B in addition to the broken relative symmetry the global $U(1)$ symmetry group is also broken, i.e. the symmetry breaking scheme is $U(1) \times SO(3)_L \times SO(3)_S \rightarrow SO(3)_J$. As a result the triads in $^3$He-B become complex. This suggests the possible consideration of the extended symmetry also in the ADW scenario. Indeed, one can see that the original fermionic action in the ADW theory is invariant under the coordinate transformation $x^\mu \rightarrow i x^\mu$, where $i$ is the imaginary unit. We called this additional symmetry the $\hat {\bf i}$-symmetry. This $Z_4$ symmetry is the discrete analogue of the  $U(1)$ symmetry in the symmetric state of liquid $^3$He. In the ADW scenario, this $\hat {\bf i}$-operation leads to the following transformation of the emerging tetrad fields: $e^a_\mu \rightarrow -i e^a_\mu$. 

The physical meaning of the spacetime coordinate transformation $x^\mu \rightarrow i x^\mu$ is discussed in Sec. \ref{imaginary}. This transformation corresponds to the transition between two equivalent descriptions of the quantum fields and gravity, Hermitian and anti-Hermitian. 

We also considered  the alternative scenario of the breaking of the $Z_4$ symmetry. 
In this scenario, the $\hat {\bf i}$-operation changes sign of the scalar curvature, ${\cal R} \rightarrow - {\cal R}$, and thus the Einstein-Hilbert action violates the $\hat {\bf i}$-symmetry. This means that in the alternative scenario of symmetry breaking, the gravitational coupling $K=1/16\pi G$ plays the role of the order parameter, with $K \rightarrow -K$ under the $\hat {\bf i}$ symmetry operation. In this scenario, the scalar field is transformed to the ghost field under ${\hat {\bf i}}$-operation; the massive particles transform to the tachyons with imaginary mass, and the de Sitter state is transformed to the anti-de Sitter state.\cite{Volovik2024f} The latter is different from the time reversal operation, which transforms the expanding de Sitter state with the Hubble parameter $H>0$ to the contracting de Sitter state with $H<0$.

The discrete $Z_4$-symmetry and its breaking can be important in cosmology. 
In particular, due to spontaneously broken discrete symmetry in the gravitational sector, gravity can be a ‘player’ in the problem of the baryon asymmetry of the Universe.\cite{Vergeles2022,Vergeles2024} Also, the breaking of discrete  symmetry leads to formation of the cosmological domain walls,\cite{Zeldovich1974,Kibble1976} see review \cite{Saikawa2017}. In Ref. \cite{Vergeles2022} the domain wall emerging due to breaking of the $Z_2$ symmetry, the $PT$-symmetry, was considered. It is the wall separating the states with $e^a_\mu$ and $-e^a_\mu$. In the case of the $Z_4$ symmetry breaking, one has in addition the H-antiH domain wall separating the quantum vacua with Hermitian and anti-Hermitian tetrads. Each of the two degenerate states can be described in the frame of the Hermitian physics by the redefinition of the $\gamma$-matrices, then its partner behind the wall is viewed as anti-Hermitians. Within the domain wall the  ${\hat {\bf i}}$-symmetry can be restored, which means that the tetrads cross the zero values, $e^a_\mu=0$. However, the symmetry of the topological objects can be also broken in their cores.\cite{SalomaaVolovik1987,Rantanen2024} The broken symmetry of the H-antiH domain wall may lead to the complex values of the tetrads within the domain wall. This would correspond to the Neel or Bloch domain walls in ferromagnets, where the magnetization does not cross zero value.

It would be interesting to extend the consideration to the extra dimensions. Diakonov suggested the $SO(16)$ symmetry group with $16\times 16$ components of the vielbein, and these 256 degrees of freedom come form the bilinear combinations of the Standard Mode fermions in four generations.\cite{Diakonov2011} 
See also the compactification of the higher-dimensional spacetime in the recent paper \cite{Toporensky2024} and references therein.

 \section{Conclusion}
 \label{ConclusionSec} 
 
  The Akama-Diakonov-Wetterich quantum gravity is symmetric under the complex coordinate transformation $x^\mu \rightarrow i x^\mu$. In this paper we discussed the physical meaning of such transformation and its physical consequences. The physical meaning becomes clear, when the imaginary unit $i$ is expressed in terms of the real valued antisymmetric matrix. Then the transformation $x^\mu \rightarrow i x^\mu$ describes the transition between the Hermitian and anti-Hermitian descriptions of the quantum fields and gravity, and corresponds to the discrete element of the $Z_4$ group. 

The spontaneous breaking of this symmetry leads to formation of the "H-anti-H" walls -- the cosmological domain walls separating the quantum vacua with Hermitian and anti-Hermitian tetrads. However, each of the two degenerate states can be described in the frame of the conventional Hermitian physics by the redefinition of the Dirac $\gamma$-matrices. 

This consideration is supported by the condensed matter analogs -- the B-phase and the planar phase of superfluid $^3$He with correspondingly massive and massless Dirac fermions.


\begin{thebibliography}{99}


\bibitem{Boyle2018}
Latham Boyle, Kieran Finn, and Neil Turok,
$CPT$-Symmetric Universe,
Phys. Rev. Lett. {\bf 121}, 251301 (2018).

\bibitem{Ryu2016}
Ching-Kai Chiu, Jeffrey C.Y. Teo, Andreas P. Schnyder, Shinsei Ryu,
Classification of topological quantum matter with symmetries,
Rev. Mod. Phys. {\bf 88}, 035005 (2016).

\bibitem{Akama1978}
K. Akama, 
An Attempt at Pregeometry: Gravity with Composite Metric,
Progress of Theoretical Physics, {\bf 60}, 1900--1909 (1978).

\bibitem{Wetterich2004}
C. Wetterich,
Gravity from spinors,
Phys. Rev. D {\bf 70}, 105004 (2004).

\bibitem{Wetterich2022}
C. Wetterich,
Pregeometry and spontaneous time-space asymmetry,
JHEP 06 (2022) 069.

\bibitem{Diakonov2011}
D. Diakonov,
Towards lattice-regularized Quantum Gravity,
arXiv:1109.0091.

\bibitem{VladimirovDiakonov2014}
A.A. Vladimirov and D. Diakonov,
Diffeomorphism-invariant lattice actions,
Physics of Particles and Nuclei {\bf 45}, 800 (2014).

\bibitem{VladimirovDiakonov2012}
A.A. Vladimirov and D. Diakonov,
Phase transitions in spinor quantum gravity on a lattice,
Phys. Rev. D {\bf 86}, 104019 (2012).

\bibitem{ObukhovHehl2012}
Y.N. Obukhov and F.W. Hehl,
Extended Einstein–Cartan theory a la Diakonov: The field equations,
Phys. Lett. B {\bf 713}, 321--325 (2012).

\bibitem{Maiezza2022}
Alessio Maiezza and Fabrizio Nesti,
Parity from gauge symmetry,
Eur. Phys. J. C  {\bf 82}, 491 (2022).

\bibitem{Vergeles2023} 
S.N. Vergeles,
Phase transition near the Big Bang in the lattice theory of gravity and some cosmological considerations,
arXiv:2301.01692 [gr-qc].

\bibitem{Volovik1990} 
G.E. Volovik, 
Superfluid $^3$He-B and gravity,
Physica B {\bf 162}, 222--230 (1990).

\bibitem{Volovik2022} 
G.E. Volovik,
Combined Lorentz symmetry: lessons from superfluid $^3$He,
J. Low Temp. Phys. {\bf 206}, 1--15 (2022),
arXiv:2011.06466.

\bibitem{Leggett1973}
A.J. Leggett, 
NMR lineshifts and spontaneously broken spin-orbit symmetry. I. General concepts, 
J. Phys. C {\bf 6}, 3187 (1973).

 \bibitem{BondarenkoZubkov2022} 
S. Bondarenko and M. A. Zubkov,
Riemann–Cartan Gravity with Dynamical Signature,
JETP Lett. {\bf 116}, 54--60, (2022).

\bibitem{Bondarenko2022} 
S. Bondarenko,
Dynamical Signature: Complex Manifolds, Gauge Fields and Non-Flat Tangent Space,
Universe {\bf 8}, 497 (2022).

\bibitem{Zlosnik2024a} 
Mehraveh Nikjoo and Tom Zlosnik,
Hamiltonian formulation of gravity as a spontaneously-broken gauge theory of the Lorentz group,
Class. Quantum Grav. {\bf  41}, 045005 (2024).

\bibitem{Zlosnik2024b} 
Priidik Gallagher, Tomi S. Koivisto, W. Ostwaldi, Luca Marzola, Ludovic Varrin and Tom Zlosnik,
Consistent first-order action functional for gauge theories,
Phys. Rev. D {\bf 109}, L061503 (2024).

\bibitem{DzyalVol1980}
I.E. Dzyaloshinskii, and G.E. Volovick, 
Poisson brackets in  condensed matter,
Ann. Phys.  {\bf 125} 67--97 (1980).

\bibitem{NissinenVolovik2019}
J. Nissinen and G.E. Volovik,
Elasticity tetrads, mixed axial-gravitational anomalies, and (3+1)-d quantum Hall effect,
Physical Review Research {\bf 1}, 023007 (2019),
arXiv:1812.03175.

\bibitem{NissinenVolovik2018}
J. Nissinen and G.E. Volovik,
Tetrads in solids: from elasticity theory to topological quantum Hall systems and Weyl fermions,
ZhETF {\bf 154},   1051--1056 (2018),
JETP {\bf 127}, 948--957 (2018),
arXiv:1803.09234.

\bibitem{KlinkhamerVolovik2019} 
F.R. Klinkhamer and G.E. Volovik,
Tetrads and $q$-theory,
Pis'ma ZhETF  {\bf 109}, 369--370 (2019),
JETP Lett. {\bf 109},  364--367 (2019),
arXiv:1812.07046.

\bibitem{Burkov2023}
Jinmin Yi, Xuzhe Ying, Lei Gioia and A.A. Burkov,
 Topological order in interacting semimetals,
Phys. Rev. B {\bf 107}, 115147 (2023).

\bibitem{Volovik2024g} 
G.E. Volovik, 
Fermionic quartet and vestigial gravity,
Pis’ma v ZhETF {\bf 119}, 317--318 (2024),
JETP Letters {\bf 119}, 330--334 (2024),
arXiv:2312.09435.

\bibitem{Volovik2022g} 
G.E. Volovik,
Dimensionless physics: continuation,
ZhETF {\bf 162},  680--685 (2022),
JETP {\bf 135}, 663--670 (2022),
arXiv:2207.05754 [gr-qc].


\bibitem{Goulart2011}
E. Goulart and Santiago Esteban Perez Bergliaffa,
Effective metric in nonlinear scalar field theories,
Phys. Rev. D {\bf 84}, 105027 (2011).

\bibitem{Yang2003}
C.N. Yang,
Thematic melodies of twentieth century theoretical physics: quantization, symmetry and phase factor,
International Journal of Modern Physics A {\bf 18}, 3263--3272 (2003).

\bibitem{VolovikZubkov2014a}
G.E. Volovik and M.A. Zubkov,
Emergent Weyl spinors in multi-fermion systems,
Nuclear Physics B {\bf 881}, 514--538  (2014),
DOI 10.1016/j.nuclphysb.2014.02.018;
arXiv:1402.5700.

\bibitem{VolovikZubkov2014b}
G.E. Volovik and M.A. Zubkov,
Emergent Weyl fermions and the origin of $i=\sqrt{-1}$ in quantum mechanics,
Pis'ma ZhETF {\bf 99},  552--557 (2014);  
JETP Lett. {\bf 99},  481--486  (2014),
arXiv:1404.4084.

\bibitem{Adler2023}
Stephen L. Adler,
Trace dynamics and its implications for my work of the last two decades,
arXiv:2307.14524 [quant-ph].

\bibitem{Adler2023b}
Stephen L. Adler,
Hidden Sector Dark Matter Realized as a Twin of the Visible Universe With Zero Higgs Vacuum Expectation,
arXiv:2308.08107.


\bibitem{Makhlin2014}
Yuriy~Makhlin, Mikhail~Silaev, and G.E. Volovik,
Topology of the planar phase of superfluid $^3$He and bulk-boundary correspondence for
three-dimensional topological superconductors,
Phys. Rev. B {\bf 89} 174502 (2014);
arXiv:1312.2677.


\bibitem{Cahill2024} 
Kevin Cahill,
Tensor gauge fields and dark matter in general relativity with fermions,
J. Phys. G: Nucl. Part. Phys. {\bf 51} 055202 (2024).

\bibitem{Volovik2024f} 
G.E. Volovik, 
Thermodynamics and decay of de Sitter vacuum,
Symmetry {\bf 16}, 763 (2024),
https://doi.org/10.3390/sym16060763

\bibitem{Vergeles2022} 
S.N. Vergeles,
Domain wall between the Dirac sea and the ‘anti-Dirac sea’,
Class. Quantum Grav. {\bf 39}, 038001 (2022).

\bibitem{Vergeles2024} 
S.N. Vergeles,
Alternative idea on the origin of the baryon asymmetry in the Universe,
unpublished.

\bibitem{Zeldovich1974} 
Y.B.  Zel'dovich, I.Y. Kobzarev and L.B. Okun, 
Cosmological Consequences of the Spontaneous Breakdown of Discrete Symmetry, 
Zh. Eksp. Teor. Fiz. {\bf 67}, 3--11 (1974), 
JETP {\bf 40}, 1--5 (1975).

\bibitem{Kibble1976} 
T.W.B. Kibble,
Topology of Cosmic Domains and Strings,
J. Phys. A {\bf 9}, 1387--1398 (1976).

\bibitem{Saikawa2017}
Ken’ichi Saikawa,
A Review of Gravitational Waves from Cosmic Domain Walls,
Universe {\bf 3}, 40 (2017). 

\bibitem{SalomaaVolovik1987}
M.M. Salomaa,  G.E. Volovik, 
Quantized vortices in superfluid $^3$He, 
Rev. Mod. Phys. {\bf 59}, 533--613 (1987).

\bibitem{Rantanen2024}
Riku Rantanen and Vladimir Eltsov,
Competition of vortex core structures in superfluid $^3$He-B,
arXiv:2406.13649.

\bibitem{Toporensky2024}
Dmitry Chirkov, Alex Giacomini, Alexey Toporensky, Petr Tretyakov,
Spontaneous symmetry breaking as a result of extra dimensions compactification,
arXiv:2407.20409.

\end{thebibliography}
\end{document}